\documentclass[aps,prl,superscriptaddress,reprint,amsmath,longbibliography]{revtex4-2}
\usepackage{graphicx}
\usepackage{enumerate}
\usepackage{refcount}
\usepackage{fmtcount}
\usepackage{amssymb}
\usepackage{hyperref}
\usepackage[usenames,dvipsnames]{color}
\usepackage[normalem]{ulem}
\usepackage{bm}
\usepackage{amsmath,amsfonts,amssymb}
\usepackage{mathrsfs}
\usepackage{bbold}

\hypersetup{pdfstartview=FitH,pdfpagemode=UseOutlines,colorlinks,
citecolor=blue,linkcolor=blue,urlcolor=blue}

\newcommand{\dg}{\dagger}
\newcommand{\pdg}{\vphantom{\dagger}}
\renewcommand{\sd}{\downarrow}
\newcommand{\su}{\uparrow}

\newcommand{\expec}[1]{\langle #1 \rangle}

\definecolor{blue(munsell)}{rgb}{0.0, 0.5, 0.69}

\begin{document}

\title{Kagome chiral spin liquid in transition metal dichalcogenide moir\'{e} bilayers}

\author{Johannes Motruk}
	\email[]{johannes.motruk@unige.ch}
    \affiliation{Department of Theoretical Physics, University of Geneva, Quai Ernest-Ansermet 24, 1205 Geneva, Switzerland}
    \author{Dario Rossi}
    \affiliation{Department of Theoretical Physics, University of Geneva, Quai Ernest-Ansermet 24, 1205 Geneva, Switzerland}
\author{Dmitry A. Abanin}
    \affiliation{Department of Theoretical Physics, University of Geneva, Quai Ernest-Ansermet 24, 1205 Geneva, Switzerland}
    \affiliation{Google Research, Mountain View, CA, USA}
\author{Louk Rademaker}
    \affiliation{Department of Theoretical Physics, University of Geneva, Quai Ernest-Ansermet 24, 1205 Geneva, Switzerland}
    \affiliation{Department of Quantum Matter Physics, University of Geneva, Quai Ernest-Ansermet 24, 1205 Geneva, Switzerland}

\date{\today}

\begin{abstract}
At $n=3/4$ filling of the moir\'e flat band, transition metal dichalcogenide moir\'e bilayers will develop kagome charge order. We derive an effective spin model for the resulting localized spins and find that its further neighbor spin interactions can be much less suppressed than the corresponding electron hopping strength. Using density matrix renormalization group simulations, we study its phase diagram and, for realistic model parameters relevant for WSe$_2$/WS$_2$, we show that this material can realize the exotic chiral spin liquid phase and the highly debated kagome spin liquid. Our work thus demonstrates that the frustration and strong interactions present in TMD heterobilayers provide an exciting platform to study spin liquid physics.
\end{abstract}

\maketitle

\textit{Introduction.---}The recent surge in moir\'e materials has vastly expanded the number of experimental platforms with strongly correlated electrons. 
While this was jumpstarted by the discovery of correlated insulating states and superconductivity in twisted bilayer graphene~\cite{Cao:2018ff,Cao:2018kn,Balents:2020.725B,Andrei:2020asg},
the strength of electron correlations in bilayers of transition metal dichalcogenide (TMD) materials exceeds those in their graphene cousins~\cite{Mak:2022review}.
Experiments in TMDs have revealed signatures of Mott insulators~\cite{Tang:2020bb,Regan:2020fk,Wang:2020jv,Li:2021cd,Ghiotto:2021qc}, the quantum anomalous Hall effect~\cite{Li:2021vy}, and -- in heterobilayers -- generalized Mott-Wigner crystals at fractional fillings~\cite{Regan:2020fk,Xu:2020dx,Huang:2021io,LiWang:2021stmWigner,Liu2021,Miao:2021exciton}.
When the electron charges are localized, only the spin degree of freedom remains, and magnetism in TMD moir\'e bilayers started to be investigated in recent experiments~\cite{Wang:2022ferromagnet,Tang2023,Anderson2023}. Heterobilayers realize an extended Hubbard model on the triangular lattice~\cite{Wu:2018ic,Zhang:2019wk,Rademaker:2021arXivSOC,Zhang:2021ix}, and consequently the localized spins are highly frustrated. This frustration might lead to a spin liquid phase, an exotic state of matter whose material realization is long sought for~\cite{Savary:2016fk,Knolle2019}.

In this Letter, we show that the generalized Mott-Wigner states at $n=\pm3/4$ filling, reported for WSe$_2$/WS$_2$ bilayers~\cite{Xu:2020dx,Huang:2021io}, can realize both a {\em chiral spin liquid}~\cite{Kalmeyer1987,Schroeter.2007} and the {\em kagome spin liquid} (KSL)~\cite{Jiang2008,Yan2011,Liao2017,He2017,Lauchli2019,Iqbal2021}.
At this particular filling, electrons are localized on an effective kagome lattice, which is known for its high degree of geometrical frustration. 
Here, we demonstrate how realistic model parameters lead to an effective spin model on this kagome lattice
and investigate the model using extensive state-of-the-art density matrix renormalization group (DMRG) simulations~\cite{White1992,McCulloch2008}.
 The tunability of TMD bilayers --  changing twist angle, gate tuning, material and dielectric environment choice, pressure, and so forth -- thus allows for a systematic pursuit of spin liquid phases~\cite{Zhou2022,Szasz2020,Kiese2022,Kuhlenkamp2022}.

\begin{figure*}
    \centering
    \includegraphics[width=\textwidth]{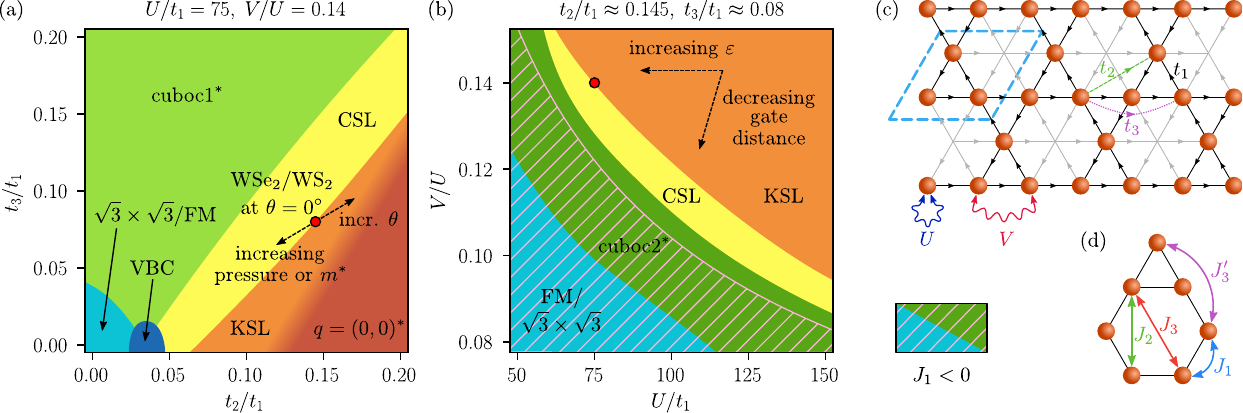}
    \caption{{\bf (a)} Ground state phase diagram of the effective spin model \eqref{eq:H_spin} for $U/t_1=75$ and $V/t_1=10.5$ -- corresponding to the interactions for $\varepsilon\approx 9.5$ in WSe$_2$/WS$_2$ at $\theta=0^\circ$ -- as a function of $t_2/t_1$ and $t_3/t_1$. The appearing phases are the chiral spin liquid (CSL), the kagome spin liquid (KSL) connected to the ground state of the nearest-neighbor Heisenberg Hamiltonian, a valence bond crystal (VBC), and cuboc1$^\ast$, $q=(0,0)^\ast$ and $\sqrt{3}\times\sqrt{3}$/ferromagnetically ordered phases. The $\sqrt{3}\times\sqrt{3}$ and ferromagnet are energetically degenerate. The red dot indicates the $t_2/t_1$ and $t_3/t_1$ values for WSe$_2$/WS$_2$ at $\theta=0^\circ$. Arrows show how this point would shift qualitatively when tuning twist angle, pressure or changing the effective mass by a different material choice. {\bf (b)} Phase diagram for $t_2/t_1 \approx 0.145$ and $t_3/t_1 \approx 0.08$ -- corresponding to WSe$_2$/WS$_2$ at $\theta=0^\circ$ -- as a function of $U/t_1$ and $V/U$. Here, the cuboc2$^\ast$ phase emerges as an additional magnetically ordered phase. The ratio of $V/U$ can be tuned by the gate distance while $U/t_1$ changes with different dielectric environment. Crucially, the system can be tuned into the CSL by mereley changing the gate distance across almost the entire range of interaction strength. The hatched area denotes the region in which $J_1$ is negative (ferromagnetic) and the red dot indicates the interaction values of panel (a). The data underlying the phase diagrams has been obtained on an infinite YC8 cylinder. {\bf (c)} Extended Hubbard model at $3/4$ filling with all charges localized on a kagome lattice. The unit cell of the charge ordering is indicated by the blue-lined box. When a spin-$\uparrow$ particle hops from a site $j$ to a site $k$ along the direction of an arrow, it picks up a phase of $\phi_{jk}$. {\bf (d)} Interactions of the resulting spin model on the kagome lattice.}
    \label{fig:Phase_Diagram}
\end{figure*}

\textit{Model.---}The moir\'{e} pattern of TMD heterobilayers is formed due to the lattice mismatch between the two layers, where the effective moir\'{e} length can be tuned by adjusting the twist angle. The interlayer band alignment ensures that the first conduction or valence flat band is completely localized in one of the layers. Based on our earlier work~\cite{Rademaker:2021arXivSOC}, we describe the resulting flat bands by a spin-orbit coupled extended Hubbard model on the triangular lattice,
\begin{eqnarray}
 	H &= 
	&\sum_{jk,\sigma} |t_{jk}| e^{-i\sigma\phi_{jk}} c^{\dg}_{j\sigma} c^{\pdg}_{k\sigma} + \text{h.c.} \nonumber \\ 
 	&&+ U \sum_j n_{j\su} n_{j\sd} + V \sum_{\expec{jk},\sigma} n_{j\sigma} n_{k \sigma}, \label{eq:H_hubb}
 \end{eqnarray}
where $\expec{jk}$ denotes nearest-neighbor sites. We include the hopping matrix elements $t_{jk}$ up to third-nearest neighbor, where $\phi_{jk}$ represent their phases induced by spin-orbit coupling.

When the nearest-neighbor repulsion is sufficiently large, a charge density wave is stabilized at commensurate fillings. In particular, at $n=\pm 3/4$ filling, the charge order forms a kagome lattice, as shown in Fig.~\ref{fig:Phase_Diagram}(c)~\cite{Pan2020,Tan2022}. In the Supplemental Material (SM)~\cite{SM,MacDondald1988,Kolley.2015}, we show, using a simple mean field theory, that the charges are almost completely localized on the kagome lattice when $V/t_1 \geq 5$.
In order to study the spin degree of freedom of these localized charges, we derive an effective spin model on the kagome lattice, starting from the extended Hubbard model of Eq.~\eqref{eq:H_hubb}~\cite{Morales2022,Morales2022arXiv}. 
In our strong coupling expansion, we keep all terms of second and third order in the hoppings, and up to fourth-order contributions proportional to the nearest-neighbor hopping $t_1$. We employ the method introduced in Ref.~\cite{Takahashi1977} and the derivation and coefficients of the model are provided in detail in the SM~\cite{SM}.%

\begin{figure}[b]
    \centering
    \includegraphics[width=\columnwidth]{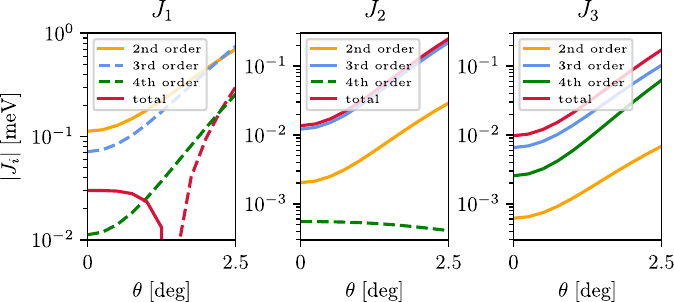}
    \caption{Contributions of the different orders of the effective Hamiltonian from our DFT estimates for WSe$_2$/WS$_2$ and dielectric screening with $\varepsilon=9.5$ as a function of twist angle. Dashed lines denote negative values.}
    \label{fig:spin_orders}
\end{figure}

The resulting spin model is given by 
\begin{eqnarray}
	 H_{\rm spin} &= & \sum_{ij}  J_{ij} \left[ S^z_i S^z_j + \cos\left(\tilde\phi_{ij}\right) (S^x_i S^x_j  +  S^y_i S^y_j ) \right.
	 \nonumber \\ 
	 &&+ \left.\sin\left(\tilde\phi_{ij}\right) (\bm{S}_i \times \bm{S}_j) \cdot \mathbf{\hat z} \right], \label{eq:H_spin}
\end{eqnarray}
where the $\tilde\phi_{ij}$ are linear combinations of the $\phi_{jk}$ phases from Eq.~\eqref{eq:H_hubb}, and we neglected very small four-spin terms. The sum over $ij$ runs over neighbors as illustrated in Fig.~\ref{fig:Phase_Diagram}(d).
The spin model of Eq.~\eqref{eq:H_spin} contains XXZ and Dzyaloshinskii-Moriya (DM) terms caused by the nonzero phases $\phi_{jk}$ in the extended Hubbard model. These phases are constrained by symmetry~\cite{Rademaker:2021arXivSOC} and translate into the phases for the spin model as follows: $\tilde \phi_1 = 4\pi/3$, $\tilde \phi_2 = 0$, and $\tilde \phi_3 = 2\pi/3$ for nearest, next-nearest and next-next-nearest neighbor couplings, respectively. This combination allows for a local three sublattice gauge transformation (a spin rotation in the $x$-$y$ plane)~\cite{SM} that brings the model into $SU(2)$-invariant form, hence, the model still exhibits a hidden $SU(2)$ symmetry~\cite{Pan2020a,Zang2021,Wietek2022}. 
As a result, the {\em structure} of the phase diagram of our kagome spin model \eqref{eq:H_spin} coincides exactly with that of an $SU(2)$-invariant $J_1$-$J_2$-$J_3$-$J_3'$ model on the kagome lattice, despite the presence of the DM interactions.
The phase diagram of this model for $J_3'=0$ has been studied previously with DMRG~\cite{Gong2015}. We remark here already that the numerical results we report are consistent with this previous work in the range of parameters studied in Ref.~\cite{Gong2015}.
Note, however, that the spin patterns in the magnetically ordered phases are changed by the gauge transformation relative to the phases of the $SU(2)$-invariant model. 

\begin{figure*}
    \centering
    \includegraphics[width=\textwidth]{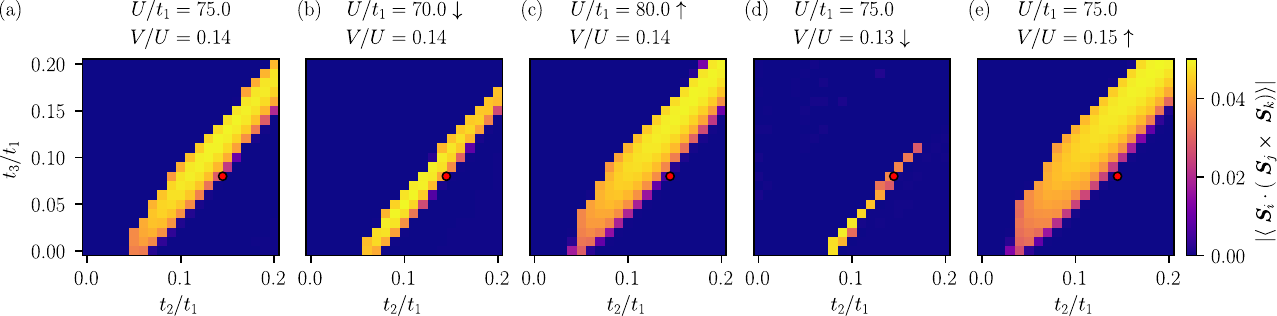}
    \caption{ CSL region as a function of $t_2/t_1$ and $t_3/t_1$ for various combinations of $U$ and $V$ from DMRG on an infinite YC8 cylinder as detected by the absolute value of the chiral order parameter $\langle \bm{S}_i \cdot (\bm{S}_j \times \bm{S}_k) \rangle$ averaged over all small nearest-neighbor triangles on the kagome lattice. {\bf (a)} $U/t_1= 75$ and $V/U=0.14$ corresponding to the phase diagram in Fig.~\ref{fig:Phase_Diagram}(a). {\bf (b)} For decreased $U/t_1=70$ with $V/U=0.14$ unchanged, the CSL region moves slightly to the lower right, but narrows. {\bf (c)} Opposite effect when increasing to $U/t_1=70$, still at $V/U=0.14$. {\bf (d)} Changing $V/U=0.13$ has a similar effect as decreasing $U$, {\bf (e)} larger $V/U=0.15$ behaves comparable to increased $U$.}
    \label{fig:chiralities}
\end{figure*}

Before presenting the numerical results, let us analyze how the spin model coefficients emerge from realistic material parameters. In Fig.~\ref{fig:spin_orders}, we show the contributions of the different orders of the expansion to $J_1$, $J_2$ and $J_3$ for WSe$_2$/WS$_2$. It is evident that the third and fourth orders are extremely important to capture the correct physics. Being ferromagnetic for $J_1$, the third and fourth orders suppress $J_1$ turning it even negative for larger twist angles (smaller interactions). In the case of $J_2$ and $J_3$, on the other hand, the spin interactions are boosted by the higher orders. In this way, $J_1$, $J_2$ and $J_3$ can be of the same order of magnitude permitting the rich phase diagrams tunable with experimental parameters we report below, despite $t_2$ and $t_3$ being an order of magnitude smaller than $t_1$. Note that this is not a sign of a breakdown of our expansion, but rather comes from the fact that the higher orders include virtual processes that do not involve intermediate states with a double occupancy and whose contribution is therefore not suppressed by factors of $1/U$.
 


\textit{Numerical results.---}To obtain the ground state phase diagram of our spin model, we perform DMRG simulations on an infinite cylinder of YC8 geometry~\cite{Yan2011,SM}. We map out the phase diagram for fixed $U/t_1=75$ and $V/U=0.14$ for varying $t_2/t_1$ and $t_3/t_1$, and for fixed $t_2/t_1\approx 0.145$ and $t_3/t_1 \approx 0.08$ for varying $U/t_1$ and $V/U$, shown in Fig.~\ref{fig:Phase_Diagram}(a) and (b), respectively. The interaction strengths of part (a) correspond to the estimates for WSe$_2$/WS$_2$ at $\theta=0^\circ$ twist angle with dielectric constant $\varepsilon \approx 9.5$. The same holds for the $t$ ratios of part (b). The derivation of these model parameters from ab initio calculations is detailed in our SM~\cite{SM}. For fixed interactions in Fig.~\ref{fig:Phase_Diagram}(a), we find two spin liquid phases, namely the CSL~\cite{He.2014,Gong2014,He.2014v0c,Gong2015,Wietek2015} and the KSL, which is connected to the nearest-neighbor Heisenberg point. For small $t_2$ and $t_3$, we observe a phase in which a ferromagnet and a $\sqrt{3}\times\sqrt{3}$ state~\cite{Harris1992,Singh1992,Sachdev1992} in the $x$-$y$ plane are the degenerate ground states due to the gauge transformation~\cite{SM}. 
In an $SU(2)$-invariant version of the model, these would be the two degenerate $\sqrt{3}\times\sqrt{3}$ states with opposite vector chirality. Next to it, we find a valence bond crystal (VBC) with spontaneous bond order. Above the diagonal, the phase diagram is dominated by the cuboc1$^\ast$ state, the gauge transformed version of the cuboc1 state, a state with finite scalar chirality~\cite{Messio:2011classical}. On the bottom right, the ground state is the $q=(0,0)^\ast$ state, the gauge transformed version of the  coplanar $q=(0,0)$ order~\cite{Zeng1990,Harris1992,Sachdev1992}. The relation between the magnetic orders, the gauge transformation and the states of the $SU(2)$-invariant Hamiltonian is further discussed in the SM~\cite{SM}.

The phase diagram in Fig.~\ref{fig:Phase_Diagram}(b), with the hopping values fixed at our estimates for WSe$_2$/WS$_2$ at $\theta=0^\circ$ similarly exhibits a finite CSL region in the center. For stronger interactions, the KSL takes over. For smaller $U$ and $V$, we find the cuboc2$^\ast$ state, the gauge transformed version of the cuboc2 magnetic order~\cite{Messio:2011classical}, and again a region with degenerate ferromagnetic and  $\sqrt{3}\times\sqrt{3}$ ground states. Most of these two phases coincide with the area in which the nearest-neighbor spin interaction turns ferromagnetic, in agreement with classical phase diagrams~\cite{Messio:2011classical}.

\textit{Chiral spin liquid.---}To identify the CSL, we primarily use the chiral order parameter (OP) $\langle \bm{S}_i \cdot (\bm{S}_j \times \bm{S}_k) \rangle$ where $i,j$ and $k$ denote the sites around a small triangle in the kagome lattice formed out of nearest-neighbor bonds. The chiral OP for the various values of $U$ and $V$ is depicted in Fig.~\ref{fig:chiralities} and clearly indicates the region of the CSL. 
We observe that the CSL region widens or narrows and shifts with changing interactions.
%
We note that both the cuboc1$^\ast$ phase as well as the $q=(0,0)^\ast$ phase can attain a nonzero chirality on the small triangles, but in a staggered pattern such that it averages out over the unit cell. 

\begin{figure}[t]
    \centering
    \includegraphics[width=\columnwidth]{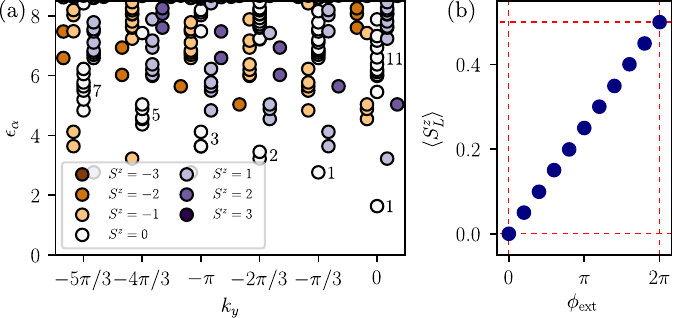}
    \caption{{\bf (a)} Momentum-resolved entanglement spectrum in the CSL phase on a YC12 cylinder. The $S^z$ sectors of the levels show the counting pattern expected from the CFT describing the edge states $(1,1,2,3,5,7,\ldots)$. {\bf (b)} Expectation value $\expec{S^z_L}$ value of the left half of the cylinder under spin flux insertion. $\Delta \expec{S^z}=1/2$ is pumped from the right to the left half of the cylinder under $2\pi$ flux insertion indicating a spin Hall conductivity of $\sigma^{\rm spin}_{xy}=1/2$. }
    \label{fig:CSL_diag}
\end{figure}

Since the chiral OP alone is not an unambiguous signature of the CSL, we also compute the momentum-resolved entanglement spectrum and the spin Hall conductivity from flux insertion. In the entanglement spectrum, a momentum $k_y$ around the cylinder and $S^z$ eigenvalue of the corresponding Schmidt state can be assigned to each  level. The chiral $SU(2)_1$ Wess-Zumino-Witten (WZW) conformal field theory describing the edge of the CSL then predicts a certain multiplet structure in each $S^z$ sector~\cite{Wen1991,Li2008,Qi2012}, which we confirm in Fig~\ref{fig:CSL_diag}(a).
We show the response of the system when threading spin flux trough the cylinder in Fig.~\ref{fig:CSL_diag}(b). We replace each term $S_j^+S_k^- \to S_j^+S_k^-e^{i\phi_{\rm ext} (y_j-y_k)/L_y}$, so that a spin up picks up a phase of $e^{i\phi_{\rm ext}}$ when going around the circumference. 
After $\phi_{\rm ext}=2\pi$ flux insertion, the expectation value of the spin in the left half of the system increases by $\expec{S^z}=1/2$ which implies a quantized spin Hall conductivity of $\sigma^{\rm spin}_{xy}=1/2$ as expected for the Kalmeyer-Laughlin CSL~\cite{Kalmeyer1987}.

\textit{Kagome spin liquid.---}The presumed ground state of the kagome spin model with only nearest-neighbor Heisenberg coupling is also a spin liquid, whose nature remains under debate~\cite{Singh2007,Jiang2008,Yan2011,Jiang2012,Liao2017,He2017,Mei.2017,Changlani2018,Lauchli2019,Jiang2019,Wietek2020,Iqbal2021}. In our $t_2$-$t_3$ phase diagram in Fig.~\ref{fig:Phase_Diagram}(a), we find a small strip of the KSL below the CSL. However, the separation between the KSL the $q=(0,0)^\ast$ state is subtle to detect. 
The spins in the $q=(0,0)^\ast$ can partly point out of the $x$-$y$ plane which happens in the region of the phase diagram that we ascribe to the $q=(0,0)^\ast$ phase. The part that we identify as the KSL has $\langle S_i^z \rangle = 0$. 
The latter region could also be a weakly ordered $q=(0,0)^\ast$ state with spins lying in the $x$-$y$ plane. However, at negative $t_3/t_1 \approx -0.032$ and $t_2/t_1\approx 0.04$, $J_2$ and $J_3$ almost vanish and we obtain a nearly only nearest-neighbor spin model. Since we find no signs of a quantum phase transition between this point and the $\langle S_i^z \rangle = 0$ region in question, we assign it to the KSL phase and take the line at which a finite $\langle S_i^z \rangle$ develops as the phase boundary.
The details of this reasoning are given in the SM~\cite{SM}. 
We emphasize that it is not within the scope of this work to give further insight into the nature of the KSL phase, but that we identify a phase with paramagnetic features which is distinct from the $q=(0,0)^\ast$ phase and adiabatically connected to the ground state of the nearest-neighbor only model. 
By this identification of phases, the entire upper right region in the $V$-$U$ phase diagram of Fig.~\ref{fig:Phase_Diagram}(b) falls into the KSL phase as well.

\begin{figure}[t]
    \centering
    \includegraphics[width=\columnwidth]{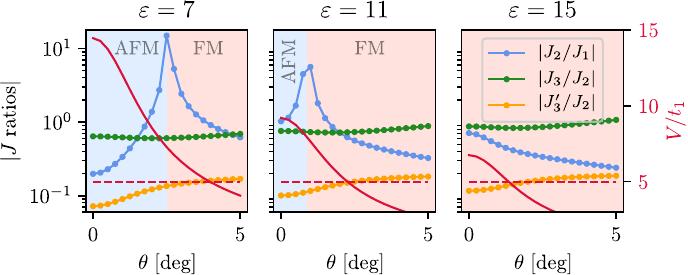}
    \caption{Absolute values of $J$ ratios from our DFT estimates for WSe$_2$/WS$_2$ as a function of twist angle for different values of $\varepsilon$. For $\varepsilon = 7$ and $11$, the ratio $J_2/J_1$ changes sign due to $J_1$ becoming negative indicated by the blue (positive) and red (negative) shading. 
    At $\varepsilon=15$, $J_1$ is negative over the entire twist angle range. $J_2$ and $J_3$ are positive everywhere while $J_3'$ is always negative. We chose the absolute values here for better presentation clarity on a log scale. We also include the ratio of $V/t_1$ (red line) and $V/t_1=5$ (red dashed line). Above this value, almost the entire particle density is localized on the kagome lattice ensuring the validity of our spin model description. }
    \label{fig:Js_eps}
\end{figure}


\textit{Experimental realization and detection.---}The red dots in our phase diagrams in Fig.~\ref{fig:Phase_Diagram} mark our estimate for the hopping and interaction values at $\varepsilon \approx 9.5$ for the first flat valence band in aligned WSe$_2$/WS$_2$~\cite{Rademaker:2021arXivSOC,SM}. We thus predict that aligned WSe$_2$/WS$_2$ falls just onto the transition line between the CSL and the KSL, suggesting a real material manifestation of these exotic spin states.
As in any TMD heterobilayer, the interaction strengths $U/t_i$ and $V/t_i$ are tunable through engineering the dielectric environment. The ratio $V/U$ can be changed by adjusting the screening length, which can be modified by the distance between the conducting gates and the bilayer. The influence of these two tuning knobs is demonstrated in Fig.~\ref{fig:Phase_Diagram}(b) which shows that the system can be driven deeper into one of the spin liquid phases.
In addition to the dielectric environment and gate distance, there are several other tuning parameters. The choice of TMD material influences the effective model -- most notably, compounds with Mo have a larger particle effective mass than We-based TMDs, which then leads to flatter bands and larger effective interactions $U/t_i$, $V/t_i$. Similarly, applying uniaxial pressure onto the bilayer increases the interlayer moiré potential, which strengthens interactions as well. We found that these two factors also lead to a slight increase in the $t_3/t_2$ ratio. On the other hand, the interaction strength can be reduced by increasing the twist angle. 

The values of the resulting spin interactions in WSe$_2$/WS$_2$ as a function of twist angle for different values of $\varepsilon$ are shown in Fig.~\ref{fig:Js_eps}. Generally, the magnitudes of the coefficients are distributed as expected with $|J_1| > |J_2| > |J_3| > |J_3'|$. 
As discussed above, $J_1$ turns negative and we obtain a ferromagnetic model for larger $\varepsilon$ and/or twist angle while $J_2$ and $J_3$ always stay positive and $J_3'$ negative. 
The relevant energy scale for the spin physics we consider is given by the nearest-neighbor exchange constant $J_1$ which is rather small due the large length scale in moir\'e systems. For the value of $U/t_1=75$ in Fig.~\ref{fig:Phase_Diagram}(a), our estimates lead to $J_1 \approx 0.03$ meV corresponding to $\approx 350$~mK which severely challenges experimental detection of the exotic spin phases. It has been proposed that magnetic order can be diagnosed by the splitting of exciton resonances~\cite{Gomez2022}. Further promising techniques include magnetic resonance force microscopy (MRFM)~\cite{Kazakova:MRFM2019}, spin-polarized scanning tunnel microscopy (STM)~\cite{Wiessendanger:2009sp-stm}, and nitrogen vacancy (NV) centers~\cite{Chatterjee2019}.
The detection of spin liquids beyond the absence of magnetic order is even more challenging. 
One possible approach is to use the optical access to the spin degree of freedom in TMDs due to spin-valley locking~\cite{Di2012,Mak2012} which may allow for the dynamical detection of the time-reversal symmetry breaking or quantized spin Hall conductivity of the CSL. Recently, magneto-optical Faraday rotation was proposed to detect the CSL in the triangular lattice Hubbard model~\cite{Banerjee2023,Szasz2020}.


\textit{Conclusion.---}We have demonstrated that a variety of magnetic phases can be realized in an effective spin model on the kagome lattice which describes TMD bilayers at a filling of 3/4 holes or electrons. In particular, the chiral spin liquid as well as the kagome spin liquid can emerge for experimentally realistic parameters, in addition to several magnetically ordered phases. 
Moreover, the tunability of TMD moir\'e systems allows for a systematic search of elusive spin liquid physics, and as such it opens up a promising new direction in the search of highly entangled quantum matter. Apart from our approach, two additional proposals for a kagome charge arrangement in twisted TMD bilayers have recently been put forward whose spin physics has yet to be investigated~\cite{Claassen2022,Reddy2023}.

The data and code used to create the reported results are available at~\cite{Motruk2023}.

\textit{Acknowledgements.---}Support by the Swiss National Science Foundation (SNSF) under grant No.~188532 and by the European Research Council (ERC) under the European Union’s Horizon 2020 research and innovation programme (grant agreement No.~864597) is gratefully acknowledged. J.~M. was supported by the SNSF Swiss Postdoctoral Fellowship grant 210478. L.~R. was funded by the SNSF via Ambizione grant 174208 and SNSF Starting Grant 211296.
DMRG simulations were performed using the \texttt{TeNPy} library~\cite{tenpy} on the Baobab and Yggdrasil HPC clusters at the University of Geneva.

\pagebreak
\widetext
\begin{center}
\textbf{\large Supplemental Material for ``Kagome chiral spin liquid in transition metal \\[0.7mm] dichalcogenide moir\'{e} bilayers''} \\[4.4mm]
Johannes Motruk,$^1$ Dario Rossi,$^1$ Dmitry A. Abanin,$^{1,2}$ and Louk Rademaker$^{1,3}$ \\[1.3mm]
{\small$^1$\textit{Department of Theoretical Physics, University of Geneva,\\[-0.5mm] Quai Ernest-Ansermet 24, 1205 Geneva, Switzerland}} \\[-0.5mm]
{\small$^2$\textit{Google Research, Mountain View, California, USA}} \\
{\small$^3$\textit{Department of Quantum Matter Physics, University of Geneva,\\[-0.5mm] Quai Ernest-Ansermet 24, 1205 Geneva, Switzerland}} \\[-0.5mm]
{\small(Dated, July 5, 2023)}
\end{center}
\vspace{3mm}
\setcounter{equation}{0}
\setcounter{figure}{0}
\setcounter{table}{0}
\setcounter{page}{1}
\makeatletter
\renewcommand{\theequation}{\arabic{equation}}
\renewcommand{\thefigure}{S\arabic{figure}}
\renewcommand{\bibnumfmt}[1]{[#1]}
\renewcommand{\citenumfont}[1]{#1}

\newcommand{\ket}[1]{\left\vert #1 \right\rangle}
\newcommand{\bra}[1]{\left\langle #1 \right\vert}

\newcommand{\nn}[1]{\left\langle #1 \right\rangle}
\newcommand{\nnn}[1]{\left\langle\left\langle #1 \right\rangle\right\rangle}
\newcommand{\nnnn}[1]{\left\langle\left\langle\left\langle #1 \right\rangle\right\rangle\right\rangle}

\thispagestyle{empty}

\section{Model parameters for WS\lowercase{e}$_2$/WS$_2$}

We derive the model for the flat bands based on the continuum model with spin-orbit coupling of Refs.~\cite{Wu:2018,Rademaker:2021}. The continuum model of a valence flat band in aligned WSe$_2$/WS$_2$ is described by the following Hamiltonian,
\begin{eqnarray}
	H&=& -\frac{\hbar^2 {\bf Q}^2}{2m^*} + V({\bf r}) 
	\nonumber
	\\
	V ({\bf r}) &=& \sum_{{\bf g}_j} V_j \exp \left[ i {\bf g}_j {\bf r} \right]
	\label{Eq:MoirePotential}
\end{eqnarray}
where $m^* = 0.36 m_e$ is the effective hole mass, ${\bf Q}$ is the momentum relative to the ${\bf K}$/${\bf K'}$ point (depending on the spin), and $V(r)$ is the moir\'e potential. The latter is characterized by the moir\'e reciprocal lattice vectors 
${\bf g}_j = \frac{4\pi}{\sqrt{3} a_M} ( -\sin \frac{2\pi (j-1)}{6},\cos \frac{2\pi (j-1)}{6} ) $ with $a_M = 7.98$ nm, and $V_1 = V_{\text{moir\'e}} \, e^{ i \psi}$ with $(V_{\text{moir\'e}},\psi) = (7.7$ meV$, 106^\circ)$ determined using ab initio density functional theory calculations~\cite{Rademaker:2021}.

\begin{figure}[b]
    \centering
    \includegraphics[width=\textwidth]{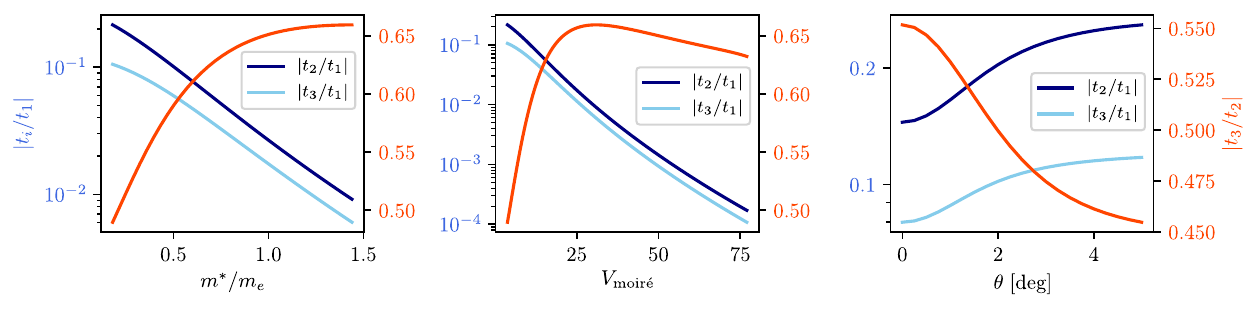}
    \caption{Behavior of the hopping parameters as a function of effective mass, moir\'e potential and twist angle.}
    \label{fig:t1_t2_t3}
\end{figure}

We then performed a Wannierization of the top valence flat band to obtain the hopping parameters $|t|$ up to third nearest neighbor. The spin-orbit coupled phases are restricted by symmetry to be $\phi_1 = \frac{2\pi}{3}$, $\phi_2 = \pi$ and $\phi_3 = \frac{\pi}{3}$. The absolute values of $|t_i|$ for aligned WSe$_2$/WS$_2$ are 
\begin{eqnarray}
    |t_1| & = & 1.81 \, \mathrm{meV},\\
    |t_2| & = & 0.26  \, \mathrm{meV},\\
    |t_3| & = & 0.14  \, \mathrm{meV}.
\end{eqnarray}
The dependence of the hopping parameters on $m^\ast$, $V_{\text{moir\'e}}$ and the twist angle is shown in Fig.~\ref{fig:t1_t2_t3}. Increasing effective mass and moir\'e potential can lead to a higher $|t_3/t_2|$ ratio (orange line) which generally favors the CSL (see Figs.~1 and 3 in the main text).
The Wannierization provides us also with a shape of the localized Wannier orbitals on the moir\'e triangular lattice. For our aligned WSe$_2$/WS$_2$, the Wannier orbital is a Gaussian centered at the W/Se region of the moir\'e unit cell with width $\sigma = 1.39$ nm.

Based on the Wannier orbital size, the onsite and nearest-neighbor repulsion can be calculated using a screened Coulomb potential
\begin{equation}
    V_C (r) = \frac{1}{4\pi \varepsilon |r|} e^{-|r| / d}
\end{equation}
where $d$ is the screening length. For an infinite screening length $d = \infty$, the resulting onsite Hubbard and nearest neighbor repulsion are expressed analytically in terms of the Wannier orbital width $\sigma$,
\begin{eqnarray}
    U & = & \frac{1}{4\pi \varepsilon \sigma} \sqrt{\frac{\pi}{2}} \\
    V & = & \frac{1}{4\pi \varepsilon \sigma} \sqrt{\frac{\pi}{2}}
        e^{- (a_M/2 \sigma)^2} \mathrm{I}_0 [ (a_M/2 \sigma)^2 ]
\end{eqnarray}
For $\varepsilon \approx 9.5$ and $d=\infty$, the values for aligned WSe$_2$/WS$_2$ are $U/t_1 = 75$ and $V/t_1 =10.5$.

\section{Mean field theory for kagome charge order}

\begin{figure}
 \includegraphics[width=0.5\textwidth]{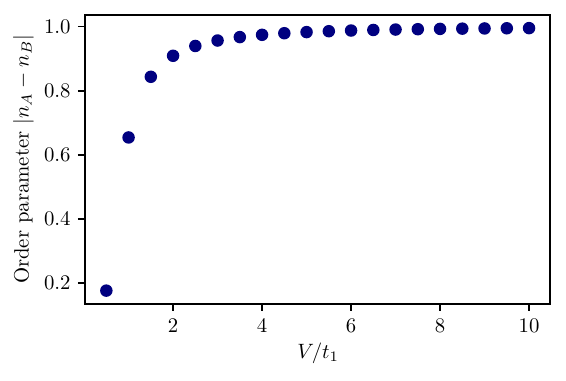}
 \caption{Kagome charge order parameter as a function of $V/t_1$ at zero temperature, based on mean field theory. Unity means that all the charge is localized on the kagome lattice.}
 \label{Fig:HFKagome}
\end{figure}

The simplest description of the kagome charge order amounts to a mean field theory for spinless electrons with nearest neighbor hopping on a triangular lattice and nearest-neighbor repulsion $V \sum_{\langle ij \rangle } n_i n_j$. The mean-field decoupling means replacing $n_i n_j \rightarrow - \langle n_i \rangle \langle n _j \rangle + n_i \langle n_j \rangle + \langle n_i \rangle n_j$. The resulting mean-field Hamiltonian is solved self-consistently for fixed particle filling $n=3/4$. The $T=0$ expectation value for the occupation difference between the kagome sites and the empty site is shown in Fig.~\ref{Fig:HFKagome}. We also find within our mean field theory that for $V/t > 0.4$ a full gap in the spectrum appears, with the charge excitation gap at large $V$ scaling as $\Delta \approx 2V - t$.

At $V/t=5$ the charge on the occupied sites exceeds $n_A = 0.98$. We choose this as threshold for charge localization, and therefore as a limit on the applicability of our strong coupling expansion of the effective spin model.

\section{Effective spin model}

\begin{figure}[b]
 \includegraphics[width=0.65\textwidth]{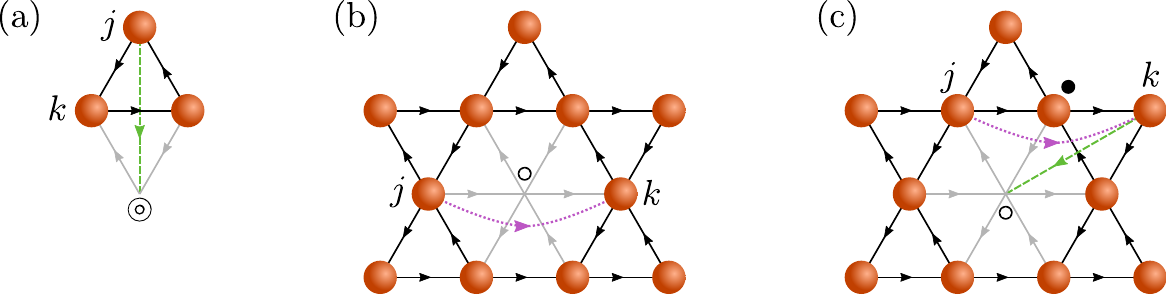}
 \caption{Notation explanation for Eq.~\eqref{eq:H3}. (a) Second line. (b) Second to last line. (c) Last line. We included the further neighbor hoppings involved in the processes creating the terms. }
 \label{fig:circles}
\end{figure}

We expand the extended Hubbard Hamiltonian from Eq.~(1) in the main text in the ratio of hoppings to interactions according to Ref.~\cite{Takahashi1977supp}. We keep all terms of second and third order in the hoppings, and all fourth order terms $\propto t_1^4$. We write the $XY$ part of the effective spin model in a more compact notation with $S^+$ and $S^-$ operators instead of $S^x$ and $S^y$. The expressions can be straightforwardly transformed into $XX/YY$ and Dzyaloshinskii-Moriya (DM) interactions. To second order, we obtain the usual antiferromagnetic expressions:
\begin{align}
 H_2 = \; & \frac{2|t_1|^2}{U-V}  \sum_{\nn{jk}} \left( 2S^z_j S^z_k + e^{-2i\phi_{jk}} S^+_j S^-_k + e^{2i\phi_{jk}} S^-_j S^+_k \right) \nonumber \\
 &+ \frac{2}{U} \Bigg[ |t_2|^2\sum_{\nnn{jk}} \left( 2S^z_j S^z_k + e^{-2i\phi_{jk}} S^+_j S^-_k + e^{2i\phi_{jk}} S^-_j S^+_k \right) \\
 & \qquad \; \; \; + |t_3|^2 \sum_{\nnnn{jk}} \left(2 S^z_j S^z_k + e^{-2i\phi_{jk}} S^+_j S^-_k + e^{2i\phi_{jk}} S^-_j S^+_k \right) \Bigg]. \nonumber
\end{align}
In the half-filled Hubbard model (one particle per site) without magnetic field, all third order terms vanish since they would break particle-hole symmetry~\cite{MacDondald1988supp}. In addition, they would generate three-spin interactions that would break time-reversal symmetry. In our case, however, there is no particle-hole symmetry and we have empty sites on the triangular lattice whose involvement can create two-spin terms at third order. They are given by
\begin{align}
  H_3 =  - \Bigg[& \frac{2(U+V)}{(U-V)V^2} |t_1|^3  \sum_{\nn{jk}} \left( 2\cos(3\phi_{jk}) S^z_j S^z_k + e^{i\phi_{jk}} S^+_j S^-_k + e^{-i\phi_{jk}} S^-_j S^+_k \right) \nonumber \\
  &+ \frac{2(U+2V)}{(U-V)V^2} \left|t_1^2 t_2\right|  \cos(\phi_{j\circledcirc}) \sum_{\nn{i j}} \left( 2 S^z_j S^z_k + e^{-2i \phi_{jk}} S^+_j S^-_k + e^{2i \phi_{jk}}  S^-_j S^+_k \right) \nonumber  \\
  &+\frac{2(U+2V)}{UV^2}\left|t_1^2 t_2\right| \sum_{\nnn{i  j}} \left( 2\cos(\phi_{jk}) S^z_j S^z_k + e^{-i \phi_{jk}} S^+_j S^-_k + e^{i  \phi_{jk}}  S^-_j S^+_k \right)  \label{eq:H3} \\
  &+ \frac{2(U+2V)}{UV^2} |t_2|^3  \sum_{\nnn{jk}} \left( 2\cos(3\phi_{jk}) S^z_j S^z_k + e^{i\phi_{jk}} S^+_j S^-_k + e^{-i\phi_{jk}} S^-_j S^+_k \right)  \nonumber \\
  &+\frac{2(U+2V)}{UV^2}\left|t_1^2 t_3\right| \sum_{\nnnn{j \circ k}} \left( 2\cos(\phi_{jk} - 2\phi_{j\circ}) S^z_j S^z_k + e^{-i ( \phi_{jk} + 2\phi_{j\circ} )} S^+_j S^-_k + e^{i ( \phi_{jk} + 2\phi_{j\circ} )}  S^-_j S^+_k \right) \nonumber \\
   &+\frac{4(U+2V)}{UV^2}\left|t_1 t_2 t_3\right| \sum_{\nnnn{j \bullet k} } \left( 2\cos(\phi_{j\circ}  + \phi_{\circ k}  -\phi_{jk}) S^z_j S^z_k + e^{-i ( \phi_{j\circ}  + \phi_{\circ k} + \phi_{jk} )} S^+_j S^-_k + e^{i ( \phi_{j\circ}  + \phi_{\circ k} + \phi_{jk}  )}  S^-_j S^+_k \right)  \Bigg].  \nonumber 
\end{align}
Here, the $\circledcirc$ denotes the empty site at the center of a hexagon which forms a $t_1^2t_2$ triangle with two nearest neighbors. The $\circ$ is the empty triangular lattice site inside a hexagon of the kagome lattice (between third nearest neighbors) and the $\bullet$ the filled site between third nearest neighbors on a line of sites. See illustrations in Fig.~\ref{fig:circles}. Note that the terms in the second to last line therefore do only act on third nearest neighbors across a hexagon and the ones in the last line only on third nearest neighbors along a line of sites.

The fourth order terms $\propto t_1^4$ read
\begin{equation}
\begin{aligned}
 H_4 = \; |t_1|^4 & \Bigg\{ \sum_{\nn{jk}} \Bigg[ \left(\frac{-42}{(U-V)^3}  + \frac{2}{(U-V)^2V} - \frac{10}{(U-V)V^2} - \frac{3}{V^3} + \frac{4}{(U+V)V^2} + \frac{16}{U(U-V)^2}  \right. \\
 & \qquad \qquad \left. - \frac{1}{(U+V)(U-V)V}  + \frac{32}{(U-V)^2(2U-V)}  + \frac{32}{(U-V)^2(2U-3V)} \right) \\
 & \qquad \quad \; \times \left( 2 S^z_j S^z_k + e^{-2i\phi_{jk}} S^+_j S^-_k + e^{2i\phi_{jk}} S^-_j S^+_k \right) \\
 & \qquad \quad \;  + \left( \frac{3}{(U-V)^3} + \frac{2}{(U-V)V^2} + \frac{4}{(U+V)V^2} + \frac{3}{2(U-V)^2V} - \frac{1}{2V^3} \right) \\
 & \qquad \quad \;  \times \left( 2 S^z_j S^z_k + e^{4i\phi_{jk}} S^+_j S^-_k + e^{-4i\phi_{jk}} S^-_j S^+_k \right) \Bigg] \\
 & \; \; \, + 2  \left( \frac{5}{(U-V)^3} - \frac{2}{U(U-V)^2} + \frac{6}{UV^2}  + \frac{1}{(U-V)^2V} - \frac{1}{V^3} - \frac{1}{(U-V)V^2} \right) \sum_{\nnn{jk}} \bm{S}_i \cdot \bm{S}_j \\
 & \; \; \, + 2  \left( \frac{2}{(U-V)^3} - \frac{1}{U(U-V)^2} \right) \sum_{\nnnn{i \bullet j}} \left( 2 S^z_j S^z_k + e^{-4i\phi_{i\bullet}} S^+_j S^-_k + e^{4i\phi_{i\bullet}} S^-_j S^+_k \right) \\
 & \; \; \, + \frac{6}{UV^2}  \sum_{\nnnn{i \circ j}} \left( 2 S^z_j S^z_k + e^{-4i\phi_{i\circ}} S^+_j S^-_k + e^{4i\phi_{i\circ}} S^-_j S^+_k \right) \\
 &  \; \; \, + 8\left( \frac{1}{(U-V)^3} - \frac{1}{(U-V)^2(2U-V)}- \frac{1}{(U-V)^2(2U-3V)}  \right) \\
 &  \; \; \, \times \sum_{jk-kl} \left( 2 S^z_j S^z_k + e^{-2i\phi_{jk}} S^+_j S^-_k + e^{2i\phi_{jk}} S^-_j S^+_k \right) \left( 2 S^z_l S^z_m + e^{-2i\phi_{kl}} S^+_l S^-_m + e^{2i\phi_{kl}} S^-_l S^+_m  \right) \Bigg\}
\end{aligned}
\label{eq:4th_order}
\end{equation}
Here, $\nnnn{i\circ j}$ denotes summation over third nearest neighbor pairs with empty triangle sites in between and $\nnnn{i\bullet j}$ over pairs with a filled site in between, as before. The fourth order terms mostly generate two-site spin operators as well, but also some four-spin terms. These are present on all bond pairs that are connected by a nearest neighbor bond, indicated by $jk-lm$. Note that they are only generated due to the presence of the nearest-neighbor interaction $V$ and vanish for $V=0$. Even for finite $V$, they give a very small contribution and we neglect them in all the numerical computations of this work. The full spin Hamiltonian we consider is given by 
\begin{equation}
 H_{\rm spin} = H_2 + H_3 + H_4',
\end{equation}
where the prime in $H_4'$ denotes $H_4$ after dropping the four-spin terms.

\begin{figure}[b]
    \centering
    \scalebox{-1}[1]{\includegraphics[width=0.4\textwidth]{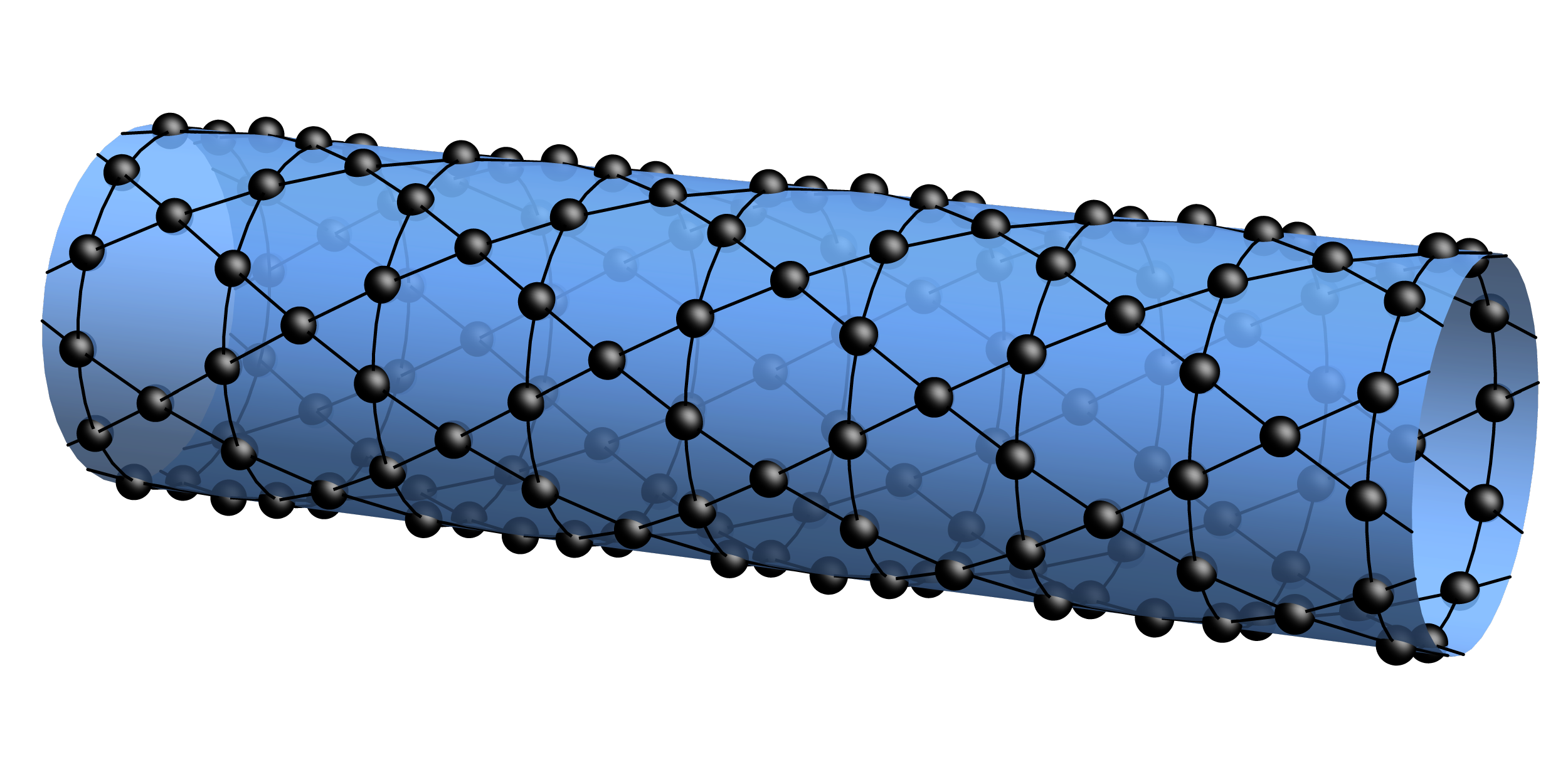}}
    \caption{YC12 cylinder geometry.}
    \label{fig:cylinder}
\end{figure}

\section{Information on DMRG simulations}

DMRG calculations are performed on infinitely long cylinders with YC8 and YC12 geometry~\cite{Yan2011supp} using the \texttt{TeNPy} package~\cite{tenpysupp}. In the YC geometry, one set of bonds of the kagome lattice is oriented along the circumference as depicted for the YC12 cylinder in Fig.~\ref{fig:cylinder}.
We use $S^z$ conservation and keep a matrix product state bond dimension of up to 3200 leading to truncation errors $\sim 10^{-5}$ for the YC8 states. All the phase diagrams in the main text are based on data of YC8 cylinders while the figures characterizing the states in Fig.~4 of the main text and Figs.~\ref{fig:ferro} to \ref{fig:KSL} in this Supplemental Material come from data on a YC12 cylinder.

\section{Gauge transformation and magnetic order of the different states}

\begin{figure}[t]
    \centering
    \includegraphics[width=0.7\textwidth]{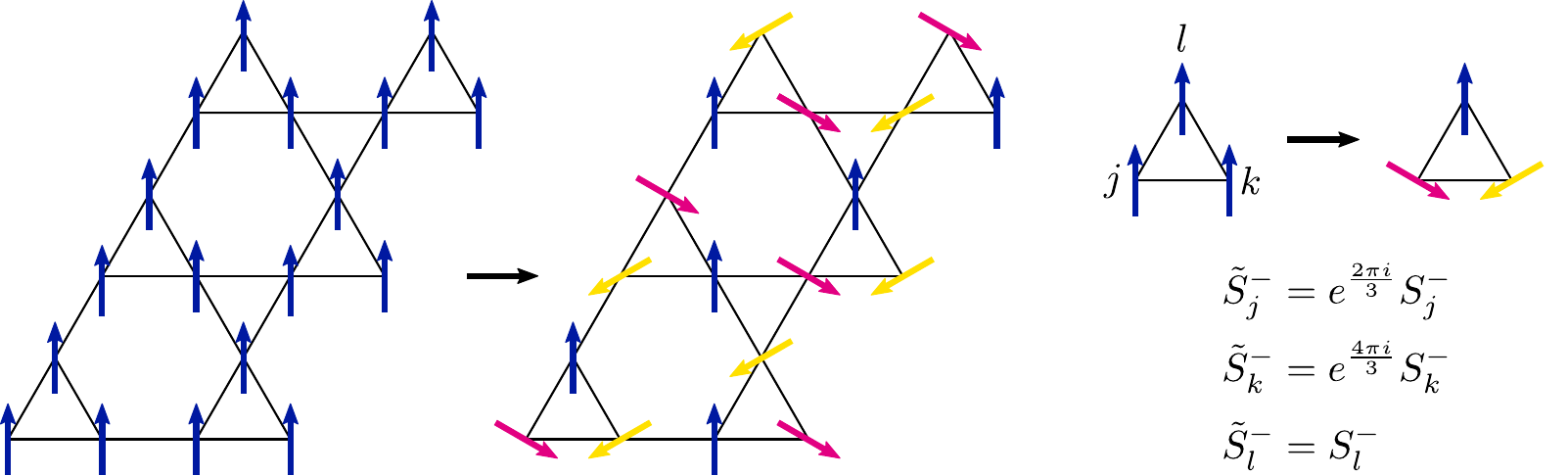}
    \caption{Local spin rotation that restores $SU(2)$ invariance of the Hamiltonian by transforming away the DM interactions. The Hamiltonian written in terms of the $\tilde S$ operators contains only Heisenberg terms. As shown here, an in-plane ferromagnet (left) develops $\sqrt{3}\times\sqrt{3}$ order (middle) under this transformation.}
    \label{fig:trafo}
\end{figure}

As mentioned in the main text, the model with $XXZ$ and DM interactions we study can be transformed into an $SU(2)$ invariant Hamiltonian by a gauge transformation, as illustrated in Fig.~\ref{fig:trafo}, which rotates the spins on different sublattices in the $xy$ plane. This transformation is related to the invariance of the spectrum of the underlying Hubbard model under the introduction of $2\pi$ flux through a triangular plaquette that has been pointed out in the literature~\cite{Zang2021supp,Wietek2022supp}. The classical regular orders that can emerge in an $O(3)$ invariant model on the kagome lattice have been classified~\cite{Messio:2011classicalsupp} and it is indeed the gauge transformed quantum versions of those which we identify in our phase diagram. For all phases, we display the spin structure factor (SF) in the extended Brillouin zone 
\begin{equation}
    S_{\mu \mu}(\bm k) = \frac{1}{N_i}\sum_{ij} \nn{S^{\mu}_i S^{\mu}_j} e^{i{\bm k}({\bm x}_j-{\bm x}_i)}
\end{equation}
with $\mu = x,y,z$. Since we conserve $S^z$ and our infinite cylinder is quasi-one-dimensional, we cannot break the remaining $U(1)$ symmetry of the model and $S_{xx}(\bm k) = S_{yy}(\bm k)$. Although we computed the phase diagram on a YC8 cylinder due to computational feasibility, the states we show here were computed on a YC12 cylinder since this geometry contains all the relevant high symmetry points in the Brillouin zone.

\subsection{Ferromagnet and $\sqrt{3}\times\sqrt{3}$ state}

\begin{figure}[t]
    \centering
    \includegraphics[width=\textwidth]{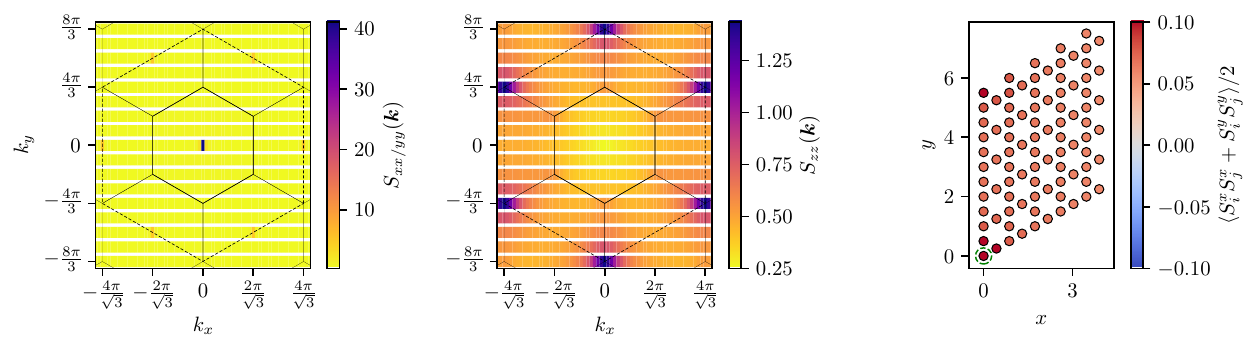}
    \caption{Structure factors and real space spin correlations in the $xy$ plane of the ferromagnetic state at $U/t_1=75, V/U=0.14, t_2/t_1 = t_3/t_1=0$. The correlations in the right panel are relative to the spin encircled in green.}
    \label{fig:ferro}
\end{figure}

Let us start with the simplest order which is the ferromagnet. Its SF and real space correlations are shown in Fig.~\ref{fig:ferro}. Since we conserve $S^z$ and work in the $m_z=0$ sector, the order has to develop in the $xy$ plane which clearly manifests itself in a peak at $\bm k=0$ in $S_{xx/yy}(\bm k)$ and the uniformly positive $xx/yy$ correlations in real space.

\begin{figure}
    \centering
    \includegraphics[width=\textwidth]{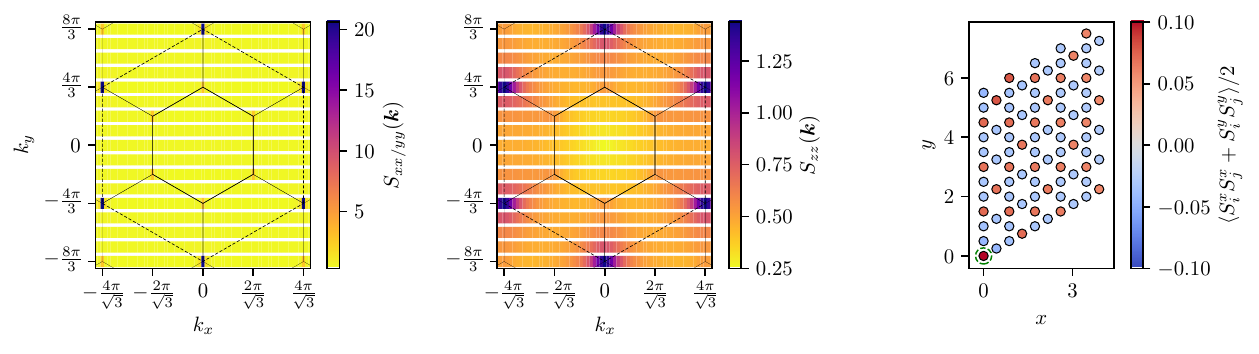}
    \caption{Structure factors and real space spin correlations in the $xy$ plane of the $\sqrt{3}\times\sqrt{3}$ state at $U/t_1=75, V/U=0.14, t_2/t_1 = t_3/t_1=0$. The $xx/yy$ SF shows clear maxima at the $\bm{K}$ points of the extended and the first Brillouin zone as expected for a $\sqrt{3}\times\sqrt{3}$ order in the $xy$ plane. The correlations in the right panel are relative to the spin encircled in green.}
    \label{fig:sqrt3}
\end{figure}

\begin{figure}
    \centering
    \includegraphics[width=0.8\textwidth]{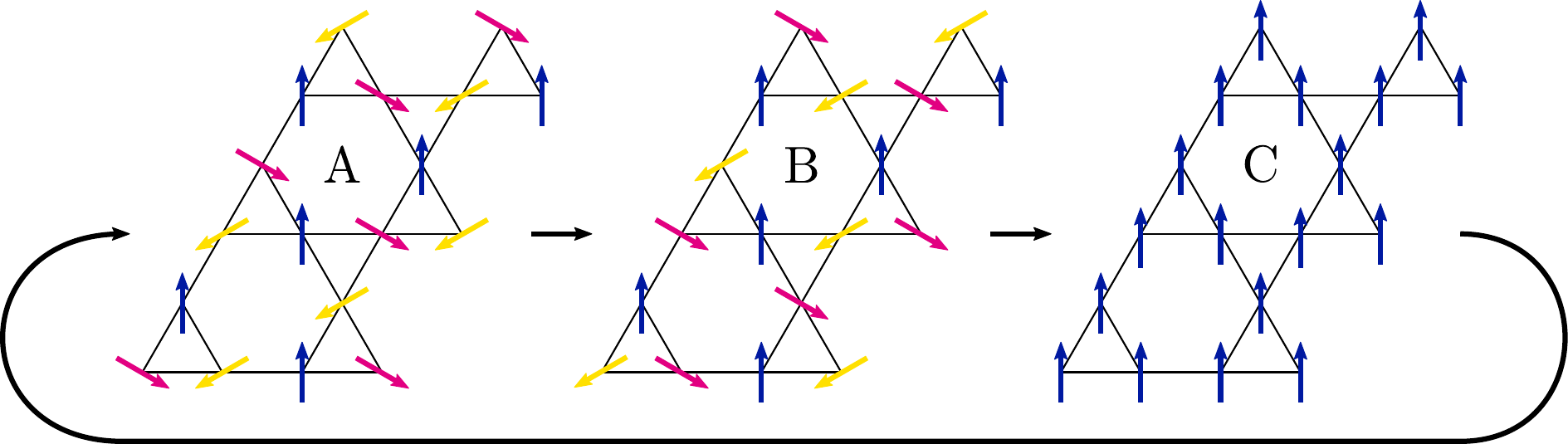}
    \caption{States related by the gauge transformation: The two vector chiralities A and B of the $\sqrt{3}\times\sqrt{3}$ state and the ferromagnet in the $xy$ plane (C).}
    \label{fig:ferro_sqrt3}
\end{figure}

The next order we consider is the coplanar $\sqrt{3}\times\sqrt{3}$ state whose SF and real space correlations are depicted in Fig.~\ref{fig:sqrt3}. This state comes in two versions of different vector chirality with $\nn{({\bm S}_i \times {\bm S}_j)\cdot \bm{\hat z}}$ positive or negative when $i$ and $j$ are adjacent sites in going around a triangle in counterclockwise direction (A and B in Fig.~\ref{fig:ferro_sqrt3}). As indicated in Fig.~\ref{fig:ferro_sqrt3}, the gauge transformation permutes these states together with the ferromagnet (C). In the $SU(2)$ invariant model, the region for small $t_2$ and $t_3$ in Fig.~1(a) of the main text would have the two different $\sqrt{3}\times\sqrt{3}$ states A and B as its degenerate ground states. However, since our spins are transformed, the ground states are B and C and the DMRG spontaneously coverges to a $\sqrt{3}\times\sqrt{3}$ or an in-plane ferromagnet in that region. With the appropriate initialization, we can reach both states at $t_2=t_3=0$ as demonstrated in Figs.~\ref{fig:ferro} and \ref{fig:sqrt3}.

\begin{figure}
    \centering
    \includegraphics[width=\textwidth]{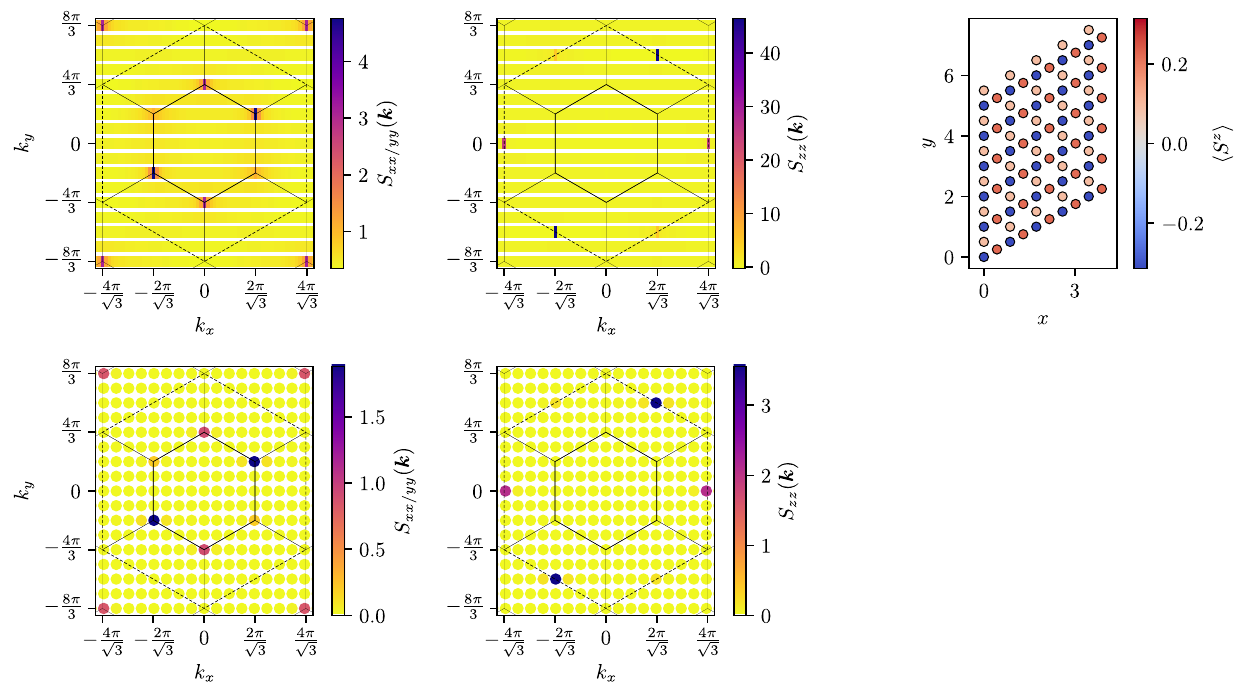}
    \caption{Structure factors and expectation values $\nn{S^z_i}$ of the $q=(0,0)^\ast$ state at $U/t_1=75, V/U=0.14, t_2/t_1 = 0.2, t_3/t_1=0$. The lower two panels show the structure factor of a classical state with $q=(0,0)$ spin pattern and applied rotations plus subsequent gauge transformation. The spins of the starting state have been rotated by $\phi = \frac{3\pi}{5} $ around the $z$ axis and $\theta = \frac \pi2$ around the $y$ axis. After that, the $xy$ part has been rotated according to the gauge transformation from Fig.~2 of the main text.}
    \label{fig:q0}
\end{figure}

\subsection{$q=(0,0)^\ast$ state}

The $q=(0,0)$ state of the $SU(2)$ invariant model is a coplanar order of spins pointing outward in a $120^\circ$ pattern on all upward or downward pointing triangles of the kagome lattice in the orientation it is drawn in Fig.~\ref{fig:ferro_sqrt3}. However, the ordering plane does not necessarily have to be the $xy$ plane since any $SU(2)$ rotated version describes the same order. We observe states whose spins are rotated out of the $xy$ plane as demonstrated in the right panel of Fig.~\ref{fig:q0}. Remember that the gauge transformation does not influence the spin value in $z$ direction. We call the gauge transformed version $q=(0,0)^\ast$ with an asterisk state to indicate the relation to the $q=(0,0)$ of the $SU(2)$ invariant model. 
If we denote the gauge transformation by an operator $G$ and an $SU(2)$ rotation by $R$ and we find a magnetically ordered ground state $\ket{\psi}$ in our model, then any state $\ket{\psi'} = G R G^{-1} \ket{\psi}$ is also a ground state.
We note that a $q=(0,0)^\ast$ state with nonzero $\nn{S^z}$ generically acquires a finite chirality since the gauge transformation rotates some of the spins out of the ordering plane. In particular, upward and downward pointing triangles develop opposite chirality which is why the average shown in Fig.~3 of the main text remains zero. In~$z$ direction, the peaks of the $q=(0,0)^\ast$ state stay at the $M$ points of the extended Brillouin zone as in the untransformed $q=(0,0)$ state,  shown in Fig.~\ref{fig:q0}. In the $xx/yy$ SF, however, they are shifted to the $\bm{K}$ points of the first Brillouin zone by the gauge transformation. We compared the SF found by DMRG to a classical one where we provide the rotation angles of the ordering plane out of the $xy$ plane and find good qualitative agreement. 

\begin{figure}
    \centering
    \includegraphics[width=\textwidth]{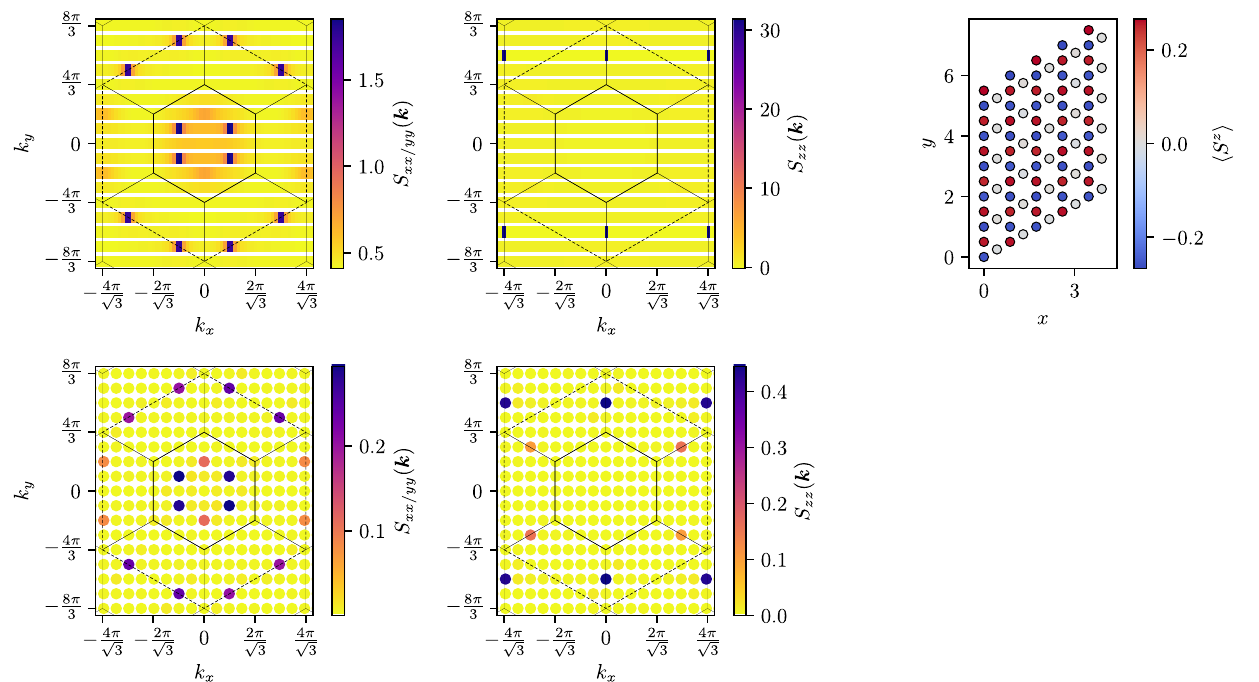}
    \caption{Structure factors and expectation values $\nn{S^z_i}$ of the cuboc1$^\ast$ state at $U/t_1=75, V/U=0.14, t_2/t_1 = 0.06, t_3/t_1=0.08$. The lower two panels show the structure factor of the classical state given in Ref.~\cite{Messio:2011classicalsupp} plus rotations and gauge transformation as in Fig.~\ref{fig:q0}. The rotation angles are $\phi = \frac{74\pi}{90} $ and $\theta = \frac{\pi}{2} $.}
    \label{fig:cuboc1}
\end{figure}

\begin{figure}
    \centering
    \includegraphics[width=\textwidth]{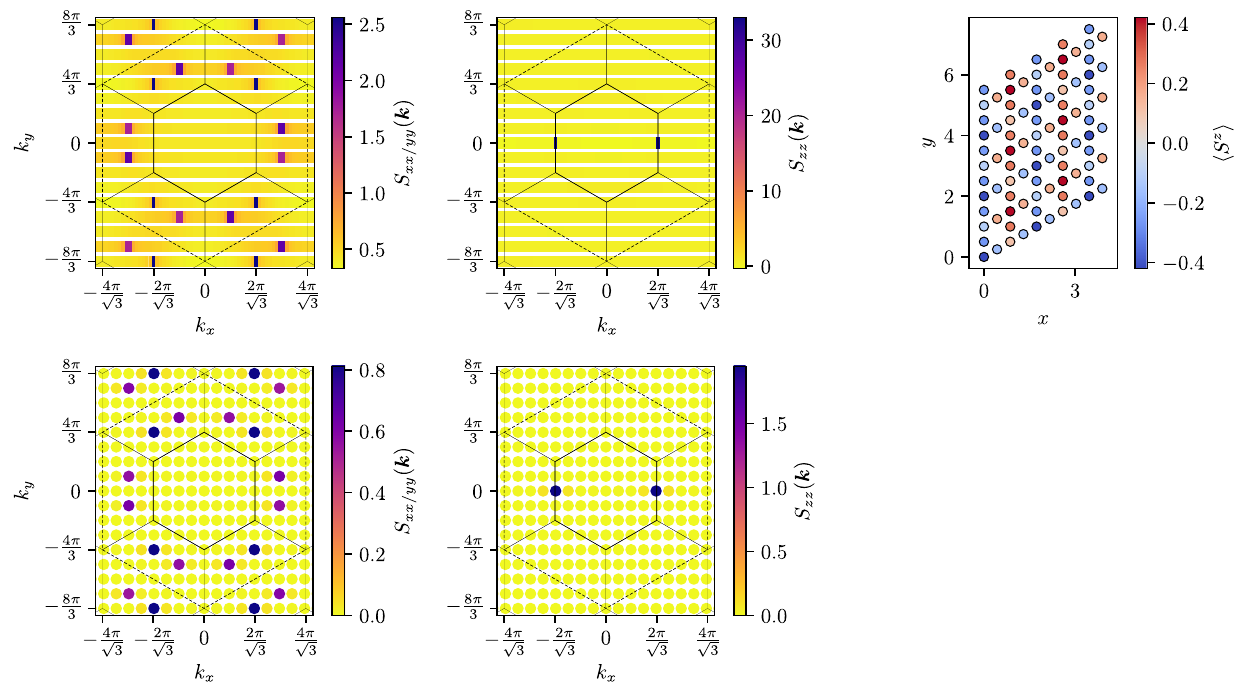}
    \caption{Structure factors and expectation values $\nn{S^z_i}$ of the cuboc2$^\ast$ state at $U/t_1=75, V/U=0.11, t_2/t_1 \approx 0.145, t_3/t_1 \approx 0.08$. The lower panels again show the rotated and gauge transformed SF of the classical state of Ref.~\cite{Messio:2011classicalsupp} with $\phi=\frac{74\pi}{90}$ and $\theta=\frac{2\pi}{3}$.}
    \label{fig:cuboc2}
\end{figure}

\subsection{Cuboc states}

The last spin rotation symmetry broken phase in the phase diagram of Fig.~1(a) of the main text is the cuboc1$^\ast$ phase. In the $SU(2)$ invariant model, the cuboc1 is a state with a 12-site unit cell in which the 12 spins point towards the corners of an eponymous cuboctahedron. Since the gauge transformation has a 9-site unit cell with $\sqrt{3}\times\sqrt{3}$ structure, the unit cell of the cuboc1$^\ast$ generally contains 144 sites and it is not very instructive to illustrate the exact spin orientations. As in the case of the $q=(0,0)^\ast$ state, we compare the structure factors of Fig.~\ref{fig:cuboc1} with a classical SF.  We therefore start from the cuboc1 pattern as classified in Ref.~\cite{Messio:2011classicalsupp}, rotate it and perform the gauge transformation. The rotation angles are given in Fig.~\ref{fig:cuboc1}.

In Fig.~\ref{fig:cuboc2}, we also provide the structure factors and $\nn{S^z}$ for the cuboc2$^\ast$ state that appears in the phase diagram of Fig.~1(b) in the main text. The same reasoning as for the cuboc1$^\ast$ applies and we again compare to the SF of a classical rotated and transformed cuboc2 state. We note that the chirality of the small triangles in the kagome lattice of both cuboc$^\ast$ states averages to zero over the unit cell which is why only the CSL displays a finite value in Fig.~3 of the main text.

\subsection{Valence bond crystal (VBC)}

\begin{figure}
    \centering
    \includegraphics[width=\textwidth]{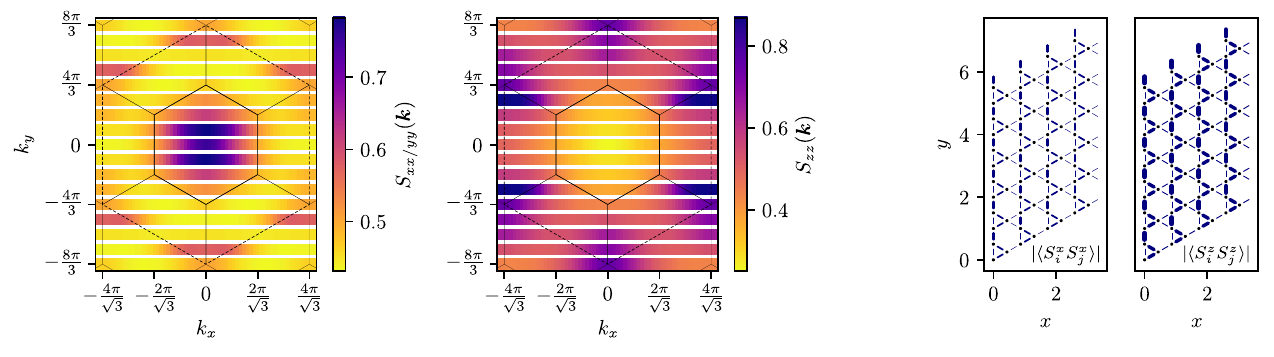}
    \caption{Structure factors and bond spin correlations $\nn{S_i S_j}$ of the valence bond crystal at $U/t_1=75, V/U=0.14, t_2/t_1 = 0.03, t_3/t_1 =0$. The thickness of the lines is proportional to the bond strength.}
    \label{fig:VBC}
\end{figure}

The valence bond crystal is a state beyond classical order that is characterized by singlet formation over certain bonds. It does not break any spin rotation symmetries, but the translation symmetry of the lattice and was also reported in the $J_1$-$J_2$-$J_3$ kagome model study of Ref.~\cite{Gong2015supp}. We show the structure factors and bond spin correlations in Fig.~\ref{fig:VBC}. No clear peaks are visible in the SF indicating the absence of spin rotation symmetry breaking. In the bond spin correlations, the pattern of strengths is consistent with the one found in Ref.~\cite{Gong2015supp}. Note that the $xx$ correlations are smaller by a factor of $1/2$ compared to the $zz$ correlations which is again a consequence of the gauge transformation. The $S^x_i$ operator of a site at which the spin operators are rotated by $2\pi/3$ expressed in the $SU(2)$ invariant $\tilde S_i$ operators reads
\begin{equation}
    S^x_i = \cos{\frac{2\pi}{3}}\tilde S_i^x - \sin{\frac{2\pi}{3}}\tilde S_i^y.
\end{equation}
Any pair of sites $\nn{ij}$ along a nearest neighbor bond has a relative $2\pi/3$ gauge rotation between them so that the generic $xx$ expectation value for such pairs is
\begin{figure}[b]
    \centering
    \includegraphics[width=\textwidth]{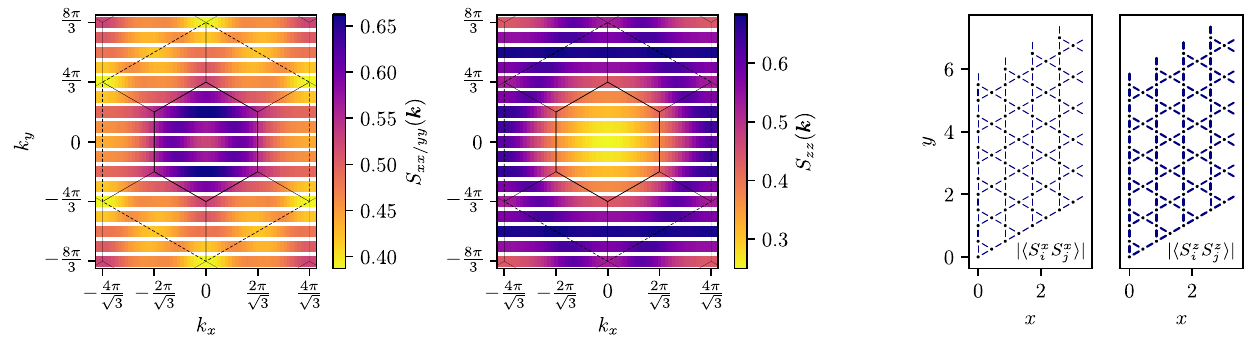}
    \caption{Structure factors and bond spin correlations $\nn{S_i S_j}$ of the chiral spin liquid at $U/t_1=75, V/U=0.14, t_2/t_1 = 0.12, t_3/t_1 =0.09$.}
    \label{fig:CSL}
\end{figure}
\begin{equation}
    \nn{S^x_i S^x_j} = \cos{\frac{2\pi}{3}}\nn{\tilde S_i^x \tilde S_j^x} - \sin{\frac{2\pi}{3}}\nn{\tilde S_i^y \tilde S_j^x}.
\end{equation}
Since $\nn{\tilde S_i^y \tilde S_j^x}$ has to be zero in an $S^z$ conserving state without spin current and taking into account that the state is $SU(2)$ invariant in terms of the $\tilde S$ variables, it follows that
\begin{equation}
    \nn{S^x_i S^x_j} = \cos{\frac{2\pi}{3}}\nn{\tilde S_i^x \tilde S_j^x} =  \nn{\tilde S_i^z \tilde S_j^z}/2 = \nn{S_i^z S_j^z}/2
\end{equation}

\begin{figure}
    \centering
    \includegraphics[width=\textwidth]{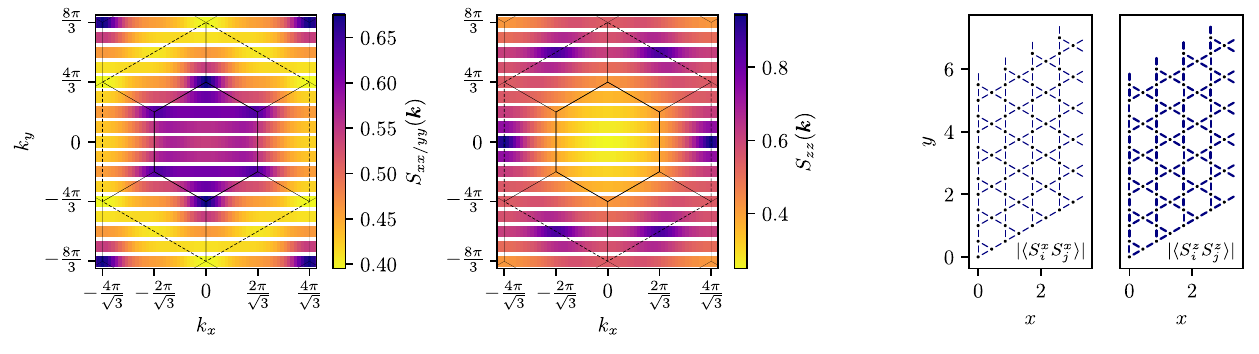}
    \caption{Structure factors and bond spin correlations $\nn{S_i S_j}$ of the kagome spin liquid at $U/t_1=125, V/U=0.14, t_2/t_1 \approx 0.145, t_3/t_1 \approx 0.08$.}
    \label{fig:KSL}
\end{figure}

\subsection{CSL and KSL}

We show the structure factors and bond spin correlations of the two spin liquid phases in Figs.~\ref{fig:CSL} and \ref{fig:KSL}. As expected for magnetically disordered phases, there are no sharp features in the SF and no pattern in the bond correlations. The SF of the KSL shows slight bumps at the same values as the $q=(0,0)^\ast$ state due to the proximity  of the two phases in parameter space.

\begin{figure}[b]
    \centering
    \includegraphics[width=0.8\textwidth]{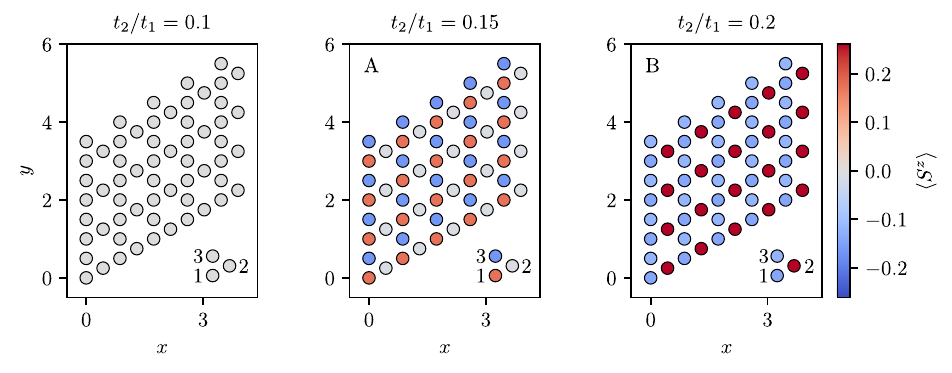}
    \caption{Expectation value $\nn{S_i^z}$ for different values of $t_2/t_1$ and $t_3/t_1=0,\; U=75$ and $V/U=0.14$ on the YC8 cylinder. The center and right panel demonstrate different orientations of the $q=(0,0)^\ast$ order. Lower right of each panel shows the three site unit cell of the kagome lattice.}
    \label{fig:diff_q0s}
\end{figure}

\section{Distinguishing KSL and $q=(0,0)^\ast$}

\begin{figure}[t]
    \centering
    \includegraphics[width=\textwidth]{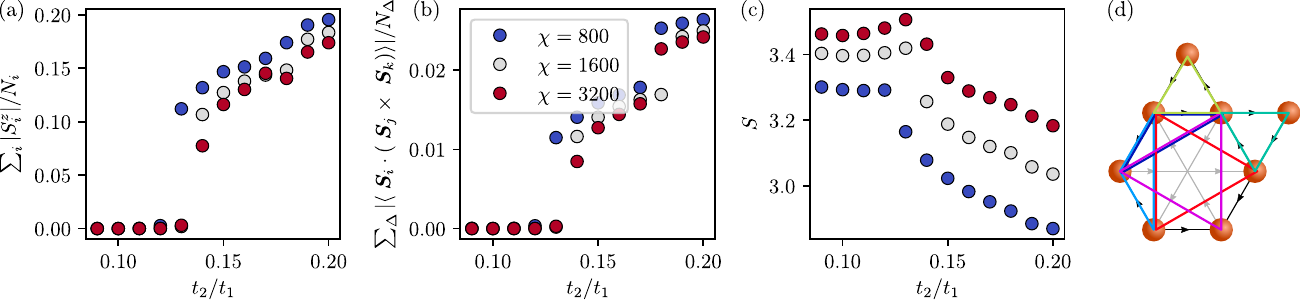}
    \caption{Different observables along the line $t_3=0$. \textbf{(a)} Average $|\langle{S^z_i}\rangle|$, \textbf{(b)} average absolute value of chirality over different triangles, \textbf{(c)} Entanglement entropy $S=-{\rm Tr} \rho_L \log \rho_L$, where $\rho_L$ is the left half-infinite cylinder. \textbf{(d)} The six different triangles per unit cell over which the absolute value of the chiral order parameter $\langle \bm{S}_i \cdot (\bm{S}_j \times \bm{S}_k) \rangle$ is averaged. In particular in (a) and (b), the transition between KSL and $q=(0,0)^\ast$ of pattern A and between pattern A and B is clearly visible at $t_2/t_1 \approx 0.13$ and $0.18$, respectively. The transition points move slightly with DMRG bond dimension $\chi$, but their locations have converged at $\chi=3200$.}
    \label{fig:cut_q0}
\end{figure}

\begin{figure}
    \centering
    \includegraphics[width=0.8\textwidth]{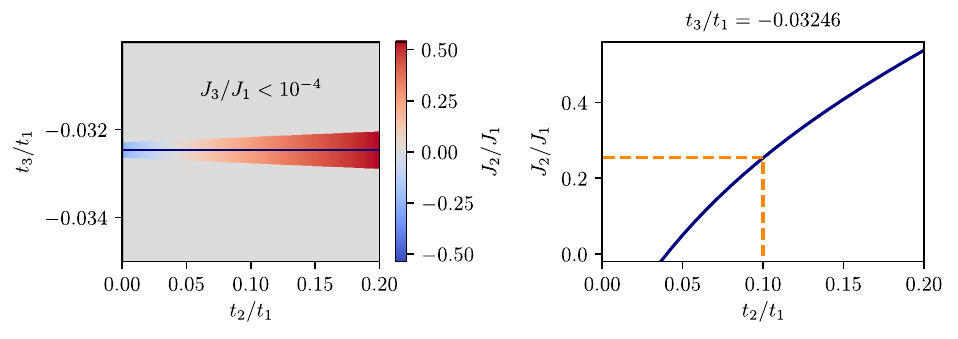}
    \caption{Left panel: The colored patch indicates the value of $J_2/J_1$ if $J_3/J_1<10^{-4}$. In the gray region, we have $J_3/J_1>10^{-4}$. Right panel: $J_2/J_1$ as a function of $t_2/t_1$ along the line of $t_3/t_1=-0.03246$ (blue line in the right panel). From DMRG data, we find the onset of finite $\nn{S^z_i}$ at $t_2/t_1 \approx 0.1$ corresponding to $J_2/J_1 \approx 0.25$.}
    \label{fig:J2_J3_zero}
\end{figure}

As explained in the previous section, the $q=(0,0)^\ast$ order can come in different orientations in our phase diagram of Fig.~1(a) of the main text. On the YC8 cylinder, we observe two orientations as identified by the $S^z$ expectation value on the individual sites. These two are illustrated in the right two panels of Fig.~\ref{fig:diff_q0s}. Orientation A (central panel) at $t_2/t_1 = 0.15$ shows $S^z$ values of $a,0$ and $-a$ at sites $1,2$ and $3$ of a small triangle whereas orientation B (right panel) at $t_2/t_1 = 0.2$ has $\left(\nn{S^z_1}, \nn{S^z_2}, \nn{S^z_3}\right) = (-b,+2b,-b)$. The KSL state at $t_2/t_1=0.1$ (left panel) has $\nn{S^z_i} = 0$ everywhere. The transitions between these regions are clearly visible in Fig.~\ref{fig:cut_q0}, where we show the average $|\langle{S^z_i}\rangle|$, the average absolute value of chirality over different triangles and the entanglement entropy
However, this could still be a state with $q=(0,0)^\ast$ order and all spins lying in the $xy$ plane. 
In order to get some more insight into this question, we investigate the coefficients of our spin model for $t_3<0$. For $t_3/t_1 = -0.03246$ and $t_2/t_1=0.04$, both $J_2/J_1$ and $J_3/J_1$ are smaller than $10^{-4}$ and $J_3'<7 \times 10^{-4}$ so that the model at this parameter point can be regarded as the nearest neighbor only model. We do not find any signs of a phase transition in any observable between this point and the $\nn{S^z_i}=0$ region in our $t_2$-$t_3$ phase diagram and therefore assign the entire region to the KSL. Furthermore, we have $J_3<10^{-4}$ on the entire line of $t_3/t_1 = -0.03246$ (see Fig.~\ref{fig:J2_J3_zero}), so that the system can be treated as the $J_1$-$J_2$ model on the kagome lattice on this parameter set. On this line, we find the onset of finite magnetization at $t_2/t_1 \approx 0.1$ which would correspond to $J_2/J_1 \approx 0.25$. This is only slightly higher than the value of $0.15$ to $0.2$ that has been provided in the DMRG literature for the transition point between the KSL and the $q=(0,0)$ in the $J_1$-$J_2$ kagome model~\cite{Gong2015supp,Kolley.2015supp}, which leads us to assign the onset of finite $\nn{S^z_i}$ as the transition to the $q=(0,0)^\ast$. An overview of the transition points in the $J_1$-$J_2$ model found by different methods is given in Table I of Ref.~\cite{Iqbal2021supp}.

\end{document}